\documentstyle[aps,amssymb,emlines,bezier,epsfig,psfig]{revtex}

\tighten
\begin{document}
\large
\title{Binding Energies and Scattering Observables
       in the $^4$He$_3$ Atomic System\thanks{LANL E-print physics/9910016}}
\author{A. K. Motovilov\thanks{On leave of absence from
         the Joint Institute  for Nuclear Research, Dubna, 141980, Russia},
    W. Sandhas}
\address{Physikalisches Institut der Universit\"at Bonn,
Endenicher Allee 11-13, D-53115 Bonn, Germany}
\author{S. A. Sofianos}
\address{Physics Department, University of South Africa,  P.O.Box
	  392, Pretoria 0003, South Africa}
\author{E. A. Kolganova}
\address{Joint Institute  for Nuclear Research, Dubna, 141980, Russia}
\date{November 8, 1999} 
\maketitle
\begin{abstract}
The  $^4$He$_3$  bound states and the scattering of a $^4$He
atom off a $^4$He dimer at ultra-low energies are investigated
using a hard-core version of the Faddeev differential equations.
Various realistic $^4$He--$^4$He interactions were employed,
amomg them the LM2M2 potential by Aziz and Slaman and the recent
TTY potential by Tang, Toennies and Yiu.  The ground state and
the excited (Efimov) state obtained are compared with other
results.  The scattering lengths and the atom-diatom phase
shifts were calculated for center of mass energies up to
$2.45$\,mK. It was found that the LM2M2 and TTY potentials,
although of quite different structure, give practically the same
bound-state and scattering results.
\end{abstract}
\section{Introduction}
\label{SIntro}
Small $^4$He clusters (in particular dimers and trimers) are of
fundamental interest in various fields of physical chemistry and
molecular physics.  Studies of these clusters represent an
important step towards  understanding  the properties of helium
liquid drops, super-fluidity in  $^4$He films, the Bose-Einstein
condensation {\it etc.} (see, for instance,
Refs.~\cite{ToenniesWinkel,RamaKrishna,LehmanScoles,GrebToeVil}).
Besides, the helium trimer is probably a unique molecular system
where a direct manifestation of the Efimov effect
~\cite{Efimov} can be observed since the binding energy
$\epsilon_d$ of the $^4$He dimer is extremely small.

The $^4$He trimer belongs to the three--body systems whose
theoretical treatment is quite difficult, first, due to its
Efimov nature and, second, because of the hard-core
properties of the inter-atomic He\,--\,He interaction
\cite{Aziz79,Aziz87,Aziz91,Tang95}.  At the same time the
problem of three helium atoms can be considered as an example of
an ideal three--body quantum problem since  $^4$He atoms are
identical neutral bosons and, thus, their handling is not complicated
by spin, isospin, or Coulomb considerations.

There is a great number of experimental and theoretical studies
of $^4$He clusters. However, most of the theoretical
investigations consist merely in computing the ground-state
energy and are based on variational methods
\cite{RickLynchDoll,Pandharip,Barnett,Lewerenz,Guardiola,Gonzalez}, on
Hyperspherical Harmonics  expansion methods  in configuration
space \cite{EsryLinGreene,Nielsen}, and on integral equations in
momentum space \cite{Nakai,Gloeckle}.  We further note that
the results of Ref.~\cite{CGM} were based on a direct solution 
of  the two-dimensional Faddeev differential equations in 
configuration space, while recently binding-energy results were 
obtained by using the three-dimensional Faddeev differential 
equations in the total-angular-momentum representation 
\cite{RoudnevYakovlev}. A qualitative treatment of the $^4$He$_3$ 
system in framework of an effective field theory is presented 
in Ref. \cite{BHvK}.  

In Refs.\cite{Gonzalez,EsryLinGreene,Gloeckle,KMS-JPB} it was
pointed out that the excited state of the $^4$He trimer is an
Efimov state \cite{Efimov}.  In these works the HFDHE2~\cite{Aziz79},
HFD-B~\cite{Aziz87}, and LM2M2~\cite{Aziz91}
versions of the $^4$He--$^4$He potentials by Aziz and co-workers
were employed.  The essential property of this state is
that it disappears when the inter-atomic potential is increased
by a factor $\lambda\sim 1.2$. And vice versa, when $\lambda$ 
slightly decreases (no more than 2\,\%), a second excited state 
appears in the trimer \cite{EsryLinGreene,Gloeckle}. It is just 
such a non-standard behavior of the excited-state energies which 
points at their Efimov nature. Another proof of such a nature of 
the excited state of the trimer, based on a scaling consideration, 
is discussed in \cite{FTDA}.  The resonance mechanism of 
formation and disappearance of the Efimov levels in the 
$^4$He$_3$ system has been studied in Refs.  
\cite{MK-FBS,KM-YAF}.

The general atom-diatom collision problem has been addressed by
various authors, and we refer the interested reader to the review
articles \cite{MichaRev81} and \cite{Kuppermann}. The collision
dynamics at thermal energies of the H$+$H$_2$ system and the
existence of resonances were discussed in \cite{KuruogluMicha}
using the Faddeev integral equations in momentum space. Finally,
the problem of existence  of $^4$He $n$-mers and their
relation to the Bose-Einstein condensation in He{\small II} was
discussed in Refs.~\cite{GhassibChester,March}. From the
experimental studies we mention those of
Refs.~\cite{DimerExp,DimerExp1,Science,ArgonExp,XenonExp,Toennies2} where
molecular clusters, consisting of a small number of noble gas
atoms, were investigated.

In contrast to the bulk of theoretical investigations devoted to the
binding energies of the $^4$He trimer, scattering processes
found comparatively little attention.  In Ref.~\cite{Nakai}
the characteristics of the He--He$_2$ scattering at zero energy
were studied, while the recombination rate of the reaction
$(1+1+1\to2+1)$ was estimated in ~\cite{Fed96}. Recently, the
phase shifts of the He--He$_2$ elastic scattering and breakup
amplitudes at ultra-low energies have  also been calculated
\cite{KMS-JPB,KMS-PRA,MSK-CPL}.

The difficulty in computing excited states and scattering
observables in the $^4$He$_3$ system is mainly due to two
reasons. First, the low energy $\epsilon_d$ of the dimer makes
it necessary to consider very large domains in configuration
space with a characteristic size of hundreds of {\AA}ngstroems.
Second, the strong repulsion of the He--He interaction at short
distances produces large numerical errors.  In the present
paper, which is an extension of our studies for $^4$He$_3$
\cite{KMS-JPB,MK-FBS,KM-YAF,KMS-PRA,MSK-CPL}, we employed the
mathematically rigorous three-body Boundary Condition Model
(BCM) of Refs. \cite{MerMot,MMYa} to the above-mentioned
problems.

As compared to \cite{KMS-JPB,MK-FBS,KM-YAF,KMS-PRA,MSK-CPL} we
employ, in the present work, the refined He--He interatomic
potentials LM2M2 by Aziz and Slaman~\cite{Aziz91}, and TTY by
Tang, Toennies, and Yiu~\cite{Tang95}.  Our numerical methods
have been substantially improved, and this allowed us to use
considerably larger grids achieving, thus, a better accuracy.
Furthermore, due to much better computing facilities more
partial waves could be taken into account.

This paper is organized as follows. In Sec. \ref{Form} we
review the three-body bound and scattering state formalism for
hard-core interactions. In Sec. \ref{Res} we describe its
application to the system of three $^4$He atoms and present our
numerical results.  Our conclusions are summarized  in Sec.
\ref{Con}. Finally in the Appendix we give details of the
potentials used.

\section{Formalism}
\label{Form}
A detailed analysis of the general boundary-value problem, the
derivation of the asymptotic boundary conditions for scattering
states and other boundary-value formulations, can be found in
Refs. \cite{MotovilovThesis,EChAYa}. In this work we employ a
hard-core version of the BCM \cite{MMYa,EfiSch} developed
in~\cite{MerMot,Vestnik} (for details see Ref. \cite{KMS-JPB}).
Therefore, in what follows we shall only  outline  the formalism
and present  its main characteristics.

In describing the three-body system we use the standard  Jacobi
coordinates  ${\bf x}_{\alpha},{\bf y}_{\alpha}$,
$\alpha=1,2,3$, expressed in terms of the position vectors of
the particles ${\bf r}_i$ and their masses ${\rm m}_i$,
\begin{eqnarray*}
\nonumber
                {\bf x}_\alpha &=&
                \left[ \frac{2{\rm m}_\beta{\rm m}_\gamma}
    {{\rm m}_\beta + {\rm m}_\gamma} \right]^{1/2}
  ({\bf  r}_\beta - {\bf  r}_\gamma)\\
%
&&\\
\nonumber
       {\bf  y}_\alpha & =&  \left[\frac
      {2{\rm m}_\alpha({\rm m}_\beta + {\rm m}_\gamma)}
       {{\rm m}_\alpha + {\rm m}_\beta + {\rm m}_\gamma}\right]^{1/2}
     \left( {\bf  r}_\alpha -\frac{{\rm m}_\beta{\bf  r}_\beta +
   {\rm m}_\gamma{\bf r}_\gamma} {{\rm m}_\beta + {\rm m}_\gamma}\right)
\end{eqnarray*}
where $(\alpha,\beta,\gamma)$ stands for a cyclic permutation of
the indices $(1,2,3)$.

In the so-called hard-core potential model one requires
that the wave function vanishes when the particles approach each
other at a certain distance $r=c$. This  requirement is
equivalent to the introduction of an infinitely strong repulsion
between the particles at distances $r\leq c$. Such a
replacement of the repulsive short-range part of the potential
by a hard-core interaction turns out to be a very efficient
way to suppress inaccuracies at short distances.
One can then show that the Faddeev components  satisfy the
following system of differential equations
\begin{equation}
\label{FaddeevEq}
\left\{
    \begin{array}{rcll}
        (-\Delta_X+V_{\alpha}-E)\Phi_{\alpha}(X)
        &=&-V_{\alpha} \displaystyle\sum
        \limits_{\beta\neq\alpha}\Phi_{\beta}(X)\,,
        \, \, & \,\, |{\bf x}_{\alpha}|>c_{\alpha}\,, \\
        (-\Delta_X-E)\Phi_{\alpha}(X)
         &=& 0\,, \,\, & \,\, |{\bf x}_{\alpha}|<c_{\alpha}\,.
     \end{array}\right.
\end{equation}
where  $X\equiv ({\bf x}_{\alpha},{\bf y}_{\alpha})$,
$\alpha=1,2,3$ and $c_\alpha$ is the hard-core radius in the
channel $\alpha$.

Outside the core the components $\Phi_\alpha$ still provide the total
wave function $\Psi$,
$$
   \Psi(X)=\sum_{\beta=1}^3 \Phi_{\beta}(X)\,,
$$
while in the interior region we have
$$
   	\sum_{\beta=1}^3 \Phi_{\beta}(X)\equiv	0.
$$
In practice, one can replace the latter strong condition by a more weak
one ~\cite{MerMot,Vestnik},
\begin{equation}
\label{SummFaddCylinder}
       \left. \displaystyle\sum_{\beta=1}^3 \Phi_\beta(X)\right |_{
       |{\bf x}_\alpha|=c_\alpha}=0, \qquad \alpha=1,2,3\,,
\end{equation}
which requires the sum of $\Phi_{\alpha}(X)$ to be zero only at
the radius $c_\alpha$.

The numerical advantage of our approach is already obvious from
the structure of Eqs.  (\ref{FaddeevEq}).  When a potential with
a strong repulsive core is replaced by the  hard-core model,
one approximates inside the core domains only the Laplacian
$\Delta_X$ instead of the sum of the Laplacian and the huge
repulsive term. In this way a much better numerical
approximation can be achieved.

In the present investigation we apply the formalism to
the $^4$He three-atomic system with total angular momentum
$L=0$. The partial-wave version
of the equations~(\ref{FaddeevEq}) for a system
of three identical bosons with  $L=0$ reads ~\cite{MF,MGL}
\begin{equation}
\label{FadPartCor}
   \left[-\displaystyle\frac{\partial^2}{\partial x^2}
            -\displaystyle\frac{\partial^2}{\partial y^2}
            +l(l+1)\left(\displaystyle\frac{1}{x^2}
            +\displaystyle\frac{1}{y^2}\right)
    -E\right]\Phi_l(x,y)=\left\{
            \begin{array}{cl} -V(x)\Psi_l(x,y), & x>c \\
                    0,                  & x<c\,.
\end{array}\right.
\end{equation}
Here, $x,y$ are the absolute values of the Jacobi variables
and $c$ is the core size which is the same for
all three two-body subsystems. The angular momentum  $l$
corresponds to a dimer subsystem and a complementary atom. For
a three-boson system in an $S$-state $l$ can only be even,
$l=0,2,4,\ldots\,.$ The potential $V(x)$ is assumed to be
central and the same for all partial waves $l$.
The function $\Psi_l(x,y)$ is related to the partial-wave
Faddeev components  $\Phi_l(x,y)$ by
\begin{equation}
\label{FTconn}
         \Psi_l(x,y)=\Phi_l(x,y) + \sum_{l'}\int_{-1}^{+1}
         {\rm d}\eta\,h_{l l'}(x,y,\eta)\,\Phi_{l'}(x',y')
\end{equation}
where
$$
        x'=\sqrt{\displaystyle\frac{1}{4}\,x^2+\displaystyle
    	\frac{3}{4}\,y^2-\displaystyle\frac{\sqrt{3}}{2}\,xy\eta}\,,
\qquad
        y'=\sqrt{\displaystyle\frac{3}{4}\,x^2+\displaystyle
   	\frac{1}{4}\,y^2+ \displaystyle\frac{\sqrt{3}}{2}\,xy\eta}\,,
$$
with $\eta=\hat{\bf x}\cdot\hat{\bf y}$. Expressions for the
kernels $h_{ll'}$ can be found in \cite{KMS-JPB,MF,MGL}.
It should be noted that these kernels depend only on the hyperangles
$$
	\theta=\arctan\frac{y}{x}\quad
	\mbox{and}\quad \theta'=\arctan\frac{y'}{x'}
$$
and not on the hyperradius
$$
   	\rho=\sqrt{x^2+y^2}=\sqrt{x'^2+y'^2}.
$$

The functions $\Phi_{l}(x,y)$ satisfy the boundary conditions
\begin{equation}
\label{BCStandard}
      	\left.\Phi_{l}(x,y)\right|_{x=0}
      	=\Phi_{l}(x,y)\left.\right|_{y=0}=0\,.
\end{equation}
The partial-wave version of the hard-core boundary
condition~(\ref{SummFaddCylinder}) reads
\begin{equation}
\label{BCCorePart}
       \Phi_{l}(c,y) + \sum_{l'}\int_{-1}^{+1}
       {\rm d}\eta\,h_{l l'}(c,y,\eta)\,\Phi_{l'}(x',y')=0\,
\end{equation}
requiring the wave function  $\Psi_{l}(x,y)$ to be zero at the
core boundary $x=c$. Furthermore, one can show that, in general, the
condition (\ref{BCCorePart}), like the
condition~(\ref{SummFaddCylinder}), causes also the wave
functions~(\ref{FTconn}) to vanish inside the core domains.  For
the  bound-state problem one requires that the functions
$\Phi_l(x,y)$ are square integrable in the quadrant $x\geq 0$,
$y\geq 0$.

The asymptotic condition for the helium trimer
scattering states reads
\begin{equation}
\label{HeBS}
    	\begin{array}{rcl}
 	\Phi_l(x,y) & = & \delta_{l0}\psi_d(x)\exp({\rm i}
 	\sqrt{E_t-\epsilon_d}\,y)
 	\left[{\rm a}_0+o\left(y^{-1/2}\right)\right] \\
        & + &
     	\displaystyle\frac{\exp({\rm i}\sqrt{E_t}\rho)}{\sqrt{\rho}}
     	\left[A_l(\theta)+o\left(\rho^{-1/2}\right)\right]
\end{array}
\end{equation}
as $\rho\rightarrow\infty$ and/or $y\rightarrow\infty$.  Here we
use the fact that the helium dimer bound state exists only  for
$l=0$.  $\epsilon_d$ stands for the dimer energy while
$\psi_d(x)$ denotes the dimer wave function which  is assumed to
be zero within the core, i.\,e., $\psi_d(x)\equiv 0$ for $x\leq
c$.

The coefficients ${\rm a}_0$ and $A_l(\theta)$ describe the
contributions of the $(2+1)$ and $(1+1+1)$ channels to
$\Phi_l$, respectively. Both the trimer binding energy $E_t$
and the difference $E_t-\epsilon_d$ in~(\ref{HeBS}) are negative
which means that for any  $\theta$ the function $\Phi_l(x,y)$
decreases exponentially as $\rho\to\infty$\,.

The asymptotic boundary condition for the partial-wave Faddeev
components of the $(2+1\rightarrow 2+1\,;\,1+1+1)$ scattering
wave function reads, as $\rho\rightarrow\infty$ and/or
$y\rightarrow\infty$,
\begin{equation}
\label{AsBCPartS}
    \begin{array}{rcl}
      \Phi_l(x,y;p) & = &
      \delta_{l0}\psi_d(x)\left\{\sin(py) + \exp({\rm i}py)
      \left[{\rm a}_0(p)+o\left(y^{-1/2}\right)\right]\right\} \\
      & + &
  \displaystyle\frac{\exp({\rm i}\sqrt{E}\rho)}{\sqrt{\rho}}
                \left[A_l(E,\theta)+o\left(\rho^{-1/2}\right)\right]
    \end{array}
\end{equation}
where $p$ is the relative momentum conjugate to the variable
$y$, $E$ is the scattering energy given by $E=\epsilon_d+p^2$,
and ${\rm a}_0(p)$ is the elastic scattering amplitude. The
functions $A_l(E,\theta)$ provide us for $E>0$ with the
corresponding partial-wave breakup amplitudes.

The helium-atom helium-dimer scattering length $\ell_{\rm sc}$ is 
given by
$$
\ell_{\rm sc}=-\displaystyle\frac{\sqrt{3}}{2}\,
\begin{array}{c}\phantom{a}\\
{\rm lim}\,\\
\mbox{\scriptsize$p\rightarrow0$}
\end{array}\,
\frac{{\rm a}_0(p)}{p}\,
$$
while the $S$-state elastic scattering phase shifts $\delta_0(p)$
are given by
\begin{equation}
\label{delta0}
    \delta_0(p)=\frac{1}{2}\,{\rm Im}\,\ln {\rm S}_0(p)
\end{equation}
where ${\rm S}_0(p)=1+2i{\rm a}_0(p)$ is the
$(2+1{\rightarrow}2+1)$ partial-wave component of the scattering
matrix.

\section{Results}
\label{Res}

We employed the Faddeev equations (\ref{FadPartCor}) and the 
hard-core boundary condition (\ref{BCCorePart}) to calculate the 
binding energies of the helium trimer and the ultra-low energy 
phase shifts of the helium atom scattered off the helium 
diatomic molecule. As He-He interaction  we used three versions 
of the semi-empirical potentials of Aziz and collaborators, 
namely HFDHE2 \cite{Aziz79}, HFD-B \cite{Aziz87}, and the newer 
version LM2M2  \cite{Aziz91}.  Further, we employed  the latest 
theoretical He--He potential TTY of Tang et al. \cite{Tang95}.  
These potentials are given in the Appendix. In our  calculations 
we used the value $\hbar^2/{\rm m}=12.12$\,K\,\AA$^2$. All the 
potentials  considered produce a weakly bound dimer state. The 
energy $\epsilon_d$ of this state together with the He--He 
atomic scattering length $\ell^{(2)}_{\rm sc}$ are given  in 
Table \ref{tableDimerLen}.  It is interesting to note that the 
latest potentials LM2M2 and TTY give practically the same 
scattering length $\ell_{\rm sc}$ and dimer energy $\epsilon_d$.

A detailed description of our numerical method has been given in 
Ref.~\cite{KMS-JPB}. Therefore, we outline here only the main 
steps of the computational scheme employed to solve the 
boundary-value problems (\ref{FadPartCor}), (\ref{BCStandard}), 
(\ref{BCCorePart}) and (\ref{HeBS}) or (\ref{AsBCPartS}).  
First, we note that the grid for the finite-difference 
approximation of the polar coordinates $\rho$ and $\theta$  is 
chosen such that the points of intersection of the arcs 
$\rho=\rho_i$, $i=1,2,\ldots, N_\rho$ and the rays 
$\theta=\theta_j$, $j=1,2,\ldots, N_\theta$ with the core 
boundary $x=c$ constitute the  knots. The value of $c$  
was fixed to be such that any further decrease of it did 
not appreciably influence the dimer binding energy $\epsilon_d$ 
and the energy of the trimer ground state $E_t^{(0)}$. In our 
previous work \cite{KMS-JPB,KMS-PRA,MSK-CPL} $c$ was chosen as 
0.7\,{\AA}. In the present work, however, we choose 
$c=1.0$\,{\AA}.  This value of $c$ provides a dimer bound state 
$\epsilon_d$ which is stable within six figures and a trimer 
ground-state energy $E_t^{(0)}$ stable within  three figures. 
The $\rho_i$ are chosen according to the formulas
\begin{eqnarray*}
	\rho_i &=&\frac{i}{N_c^{(\rho)}+1}\, c,
	\qquad i=1,2,\ldots,N_c^{(\rho)},\\
  	\rho_{i+N_c^{(\rho)}}& = & \sqrt{c^2 + y_i^2}, \qquad
	     	i=1,2,\ldots,N_\rho-N_c^{(\rho)},
\end{eqnarray*}
where $N_c^{(\rho)}$ stands for the number
of arcs inside the domain $\rho<c$ and
$$
		y_i = f(\tau_i)\sqrt{\rho^2_{N_\rho}-c^2}, \qquad
		\tau_i = \frac{i}{N_\rho-N_c^{(\rho)}}.
$$
The nonlinear monotonously increasing function  $f(\tau)$,
$0\leq\tau\leq 1$, satisfying the conditions $f(0)=0$ and
$f(1)=1$, is chosen according to
$$
		f(\tau)=\left\{
		\begin{array}{lcl} \alpha_0\tau & , & \tau\in[0,\tau_0]\\
		\alpha_1\tau+\tau^\nu &,& \tau\in(\tau_0,1]
		\end{array}
		\right..
$$
The values of $\alpha_0$, $\alpha_0\geq 0,$ and $\alpha_1$,
$\alpha_1\geq 0,$ are determined via $\tau_0$ and $\nu$ from the
continuity condition for $f(\tau)$ and its derivative at the
point $\tau_0$. In the present investigation we took values of
$\tau_0$ within 0.15 and 0.2.  The value of the power $\nu$
depends on the cutoff radius $\rho_{\rm
max}=\rho_{N_\rho}=200$--1000\,{\AA}, its range being
within 3.4 and 4 in the present calculations.

The knots $\theta_j$ at $j=1,2,\ldots,N_\rho-N_c^{(\rho)}$ are
taken according to $\theta_j={\rm arctan}(y_j/c)$ with the
remaining knots $\theta_j$,
$j=N_\rho-N_c^{(\rho)}+1,\ldots,N_\theta,$ being chosen
equidistantly.  Such a choice is required by the need of having
a higher density of points in the domain where the functions
$\Phi_l(x,y;z)$ change most rapidly, i.\,e., for small values of
$\rho$ and/or $x$. In this work, we used grids of the dimension
\mbox{$N_\theta=N_\rho=$\,500--800} while the above number
$N_c^{(\rho)}$ and the number $N_\theta-(N_\rho-N_c^{(\rho)})$
of knots in $\theta$ lying in the last arc inside the core
domain was chosen equal to 2--5.

Since we consider identical bosons only the components $\Phi_l$ 
corresponding to even $l$ differ from zero.  Thus, the number of 
equations to be solved is $N_{\rm e}=l_{\rm max}/2+1$ where 
$l_{\rm max}$ is the maximal even partial wave.  The 
finite-difference approximation of the $N_{\rm e}$ equations 
(\ref{FadPartCor}) reduces the problem to a system of $N_{\rm 
e}{N_\theta}N_\rho$ linear algebraic equations.  The 
finite-difference equations corresponding to the arc $i=N_\rho$ 
include initially the values of the unknown functions 
$\Phi_l(x,y;z)$ from the arc $i=N_\rho+1$.  To eliminate them, 
we express these values through the values of $\Phi_l(x,y;z)$ on 
the arcs $i=N_\rho$ and $i=N_\rho-1$ by using the asymptotic 
formulas~(\ref{HeBS}) or (\ref{AsBCPartS}) in the manner 
described in the final part of Appendix A of 
Ref.~\cite{KMS-JPB}. In \cite{KMS-JPB}, however, this approach 
was used for computing the binding energies only, while in the 
present work this method is extended also to the scattering 
problem. The matrix of the resulting system of equations has a 
block three-diagonal form.  Every block has the dimension 
$N_{\rm e}N_\theta\times N_{\rm e}N_\theta$ and consists of the 
coefficients standing at unknown values of the Faddeev 
components in the grid knots belonging to a certain arc 
$\rho=\rho_i$.  The main diagonal of the matrix consists of 
$N_\rho$ such blocks.

In this work we solve the block three-diagonal algebraic system 
on the basis of the matrix sweep method~\cite{Samarsky}.  The 
use of this method makes it possible to avoid writing the matrix 
on the hard drive of the computer. Besides, the matrix sweep 
method reduces the computer time required by almost one order of 
magnitude as compared to \cite{KMS-JPB,KMS-PRA,MSK-CPL}.

Our results for the trimer ground-state energy $E_t^{(0)}$, as 
well as the results obtained by  other authors, are presented in 
Table \ref{tableTrimerGS}. It should be noted that most of the 
contribution to the ground-state energy stems from the $l=0$ and 
$l=2$ partial-wave components, the latter being slightly more 
than 30\,\%, and is approximately the same for all potentials 
used.  The contribution from the $l=4$ partial wave is of the 
order of 3-4\,\% (cf.~\cite{CGM}).

It is  well known that the excited state of the $^4$He trimer is 
an Efimov state 
\cite{EsryLinGreene,Gloeckle,KMS-JPB,MK-FBS,KM-YAF}.  The 
results obtained for its energy $E_t^{(1)}$, as well as the 
results found in the literature, are presented in Table 
\ref{tableTrimerExcS}. To illustrate the convergence of our 
results we show in Table \ref{tableExcGrids} the dependence of 
$E_t^{(1)}$ on the grid parameters using the TTY potential. It 
is seen that the $l=0$ partial-wave component contributes about 
71\,\% to the excited-state binding energy. The contribution to 
$E_t^{(1)}$ from the $l=2$ component is about 25--26\,\% and 
from $l=4$ within 3--4\,\%.  These values are similar to the 
ones for the ground state.

Besides the binding-energy calculations, we  also performed
calculations for a helium atom scattered  off a helium dimer for
$L=0$. For this we used the asymptotic boundary conditions
(\ref{AsBCPartS}).  The scattering lengths of the
collision of the He atom on the He dimer obtained for the HFD-B,
LM2M2 and TTY potentials are presented in Table
\ref{tableTrimerLen}.  As compared to~\cite{KMS-JPB} the present
calcualtion is essentially improved (the result $\ell_{\rm
sc}=145{\pm}5$\,{\AA} for HFD-B with $l_{\rm max}=2$ was
obtained in \cite{KMS-JPB} with a much smaller grid).  Within
the accuracy of our calculations, the scattering lengths
provided by the LM2M2  and TTY potentials, like the energies of
the excited state, are exactly the same. This is of no
surprise since the two potentials produce practically the same
two-body binding energies and scattering lengths.

\begin{figure}
\centering
\epsfig{file=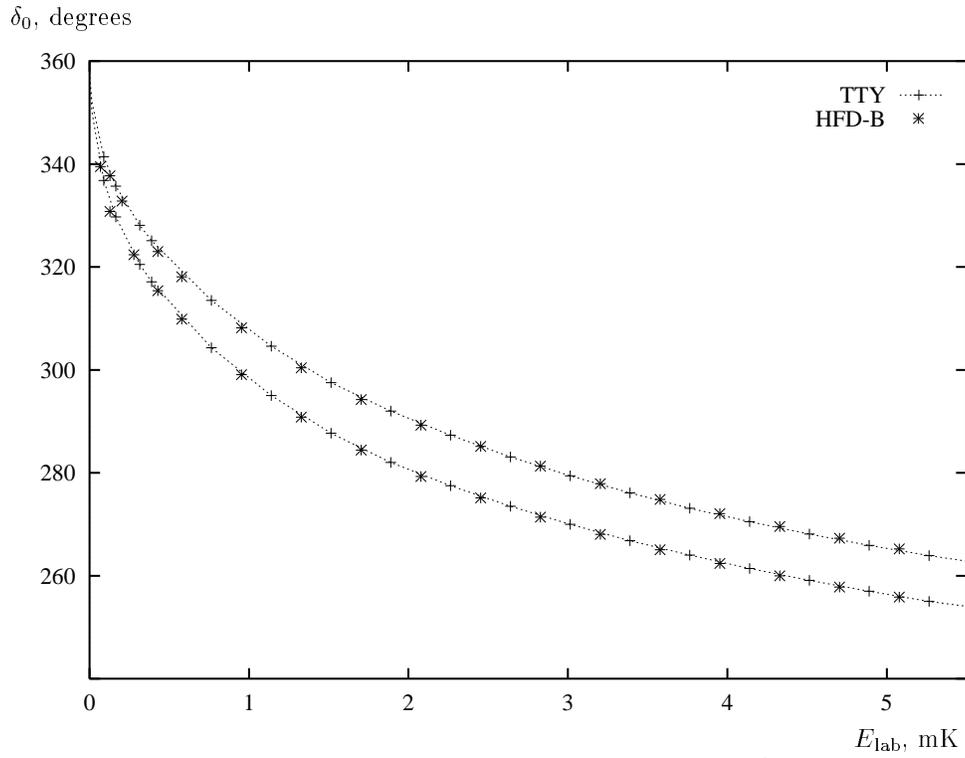,height=10cm}
\caption{S-wave helium atom -- helium dimer scattering phase
shifts $\delta_0(E_{\rm lab})$, $E_{\rm lab}=\frac{3}{2}(E+|\epsilon_d|)$,
for the HFD-B and TTY $^4$He--$^4$He
potentials. The lower curve corresponds to the case where
$l_{\rm max}=0$ while for the upper $l_{\rm max}=2$. }
\label{Fig-phases-black}
\end{figure}

The phase-shift results obtained for the HFD-B, LM2M2
and TTY potentials are given in Tables \ref{tablePhaseHFDB},
\ref{tablePhaseLM2M2}, and \ref{tablePhaseTTY}.  For the HFD-B 
and TTY potentials they are plotted in 
Fig.~\ref{Fig-phases-black}. Note that for the phase shifts 
we use the normalization required by the Levinson 
theorem \cite{Levinson}, $\delta_L(0)-\delta_L(\infty)=n\pi$, 
where $n$ is the number of the trimer bound states.

Incident energies below and above the breakup threshold were 
considered, i.\,e., for the $(2+1\longrightarrow 2+1)$ and 
$(2+1\longrightarrow 1+1+1)$ processes. It was found that after 
transformation to the laboratory system the phases 
$\delta_0^{(l_{\rm max})}$ for the potentials HFD-B, LM2M2 and 
TTY for different values of $l_{\rm max}$ are practically the 
same, especially those for LM2M2 and TTY. The difference between 
the phase shifts $\delta_0^{(2)}$  and $\delta_0^{(4)}$ is only 
about 0.5\,\%.

It is interesting to compare the values obtained for
the He\,--\,He$_2$ scattering lengths $\ell_{\rm sc}$ with
the corresponding inverse wave numbers $\varkappa^{-1}$ for the trimer
excited-state energies. The values of $\varkappa^{-1}$ where
$
   \varkappa=2\sqrt{(\epsilon_d-E_t^{(1)})/3}\,,
$
with both the $E_t^{(1)}$ and $\epsilon_d$ being given in 
{\AA}$^{-2}$, are also presented in Table \ref{tableTrimerLen}.  
It is seen that these numbers are about 1.3--1.7 times smaller 
than the respective $^4$He-atom $^4$He-dimer scattering lengths.  
The situation differs completely from the $^4$He two-atomic 
scattering problem, where the inverse wave numbers 
$\bigl(\varkappa^{(2)}\bigr)^{-1}=|\epsilon_d|^{-1/2}$ are in a 
rather good agreement with the $^4$He--$^4$He scattering lengths 
(see Table \ref{tableDimerLen}).  Such significant differences 
between $\ell_{\rm sc}$ and $\varkappa^{-1}$ in the case of the 
$^4$He three-atomic system can be attributed to the Efimov 
nature of the excited state of the trimer, which implies that 
the effective range $r_0$ for the interaction between the $^4$He 
atom and the $^4$He dimer is very large as compared to the 
$^4$He diatomic problem.

\section{Conclusions}
\label{Con}

In this paper we employed a hard-core variant of the Boundary 
Condition Model which, unlike some competing methods, is exact 
and ideally suited for three-body calculations  with two--body 
interactions characterized by a highly repulsive core. The 
method is relevant not only for bound states but also for 
scattering processes. There is, however, a price to be paid for 
the exact treatment of the system. Higher partial waves, beyond 
$l_{\rm max}=4$, were hard to be incorporated within the 
computing facilities  at our disposal.

The results for the ground-state energy of the $^4$He trimer 
obtained for all four $^4$He--$^4$He potentials considered 
compare favorably with alternative results in the literature.  
Furthermore, the treatment of the excited state, interpreted as 
an Efimov state, clearly demonstrates the reliability of our 
method in three-body bound state calculations with hard-core 
potentials. In addition to the binding energy, the formalism has 
been successfully used to calculate scattering lengths and  
ultra-low-energy phase shifts of the $^4$He atom scattered off 
the $^4$He dimer.

In general the hard-core inter-atomic potential together with 
other characteristics of the system,  makes calculations 
extremely tedious and numerically unstable.  This is not the 
case in our formalism where the hard core is taken into account 
from the very beginning in a mathematically rigorous way. Thus, 
the formalism  paves the way to study various ultra-cold 
three-atom systems, and to calculate important quantities as 
cross-sections, recombination rates, {\it etc.}


\bigskip
\acknowledgements
The authors are grateful to Prof.~V.\,B.\,Belyaev and 
Prof.~H.\,Toki for help and assistance in performing the 
calculations at the supercomputer of the Research Center for 
Nuclear Physics of the Osaka University, Japan. One of the 
authors (W.\,S.) would like to thank J.\,P.\,Toennies for very 
interesting discussions stimulating this investigation.  
Financial  support by the Deutsche Forschungsgemeinschaft, the 
Russian Foundation for Basic Research, and the National Research 
Foundation of South Africa, is gratefully acknowledged.

\appendix
\section*{The Potentials Used}
The general structure of the realistic semi-empirical potentials
HFDHE2~\cite{Aziz79} and HFD-B~\cite{Aziz87} developed by Aziz
{\it et al.} is
\begin{equation}
\label{HFD}
  	V(x)=\varepsilon\,V_b(\zeta)
\end{equation}
where  $\zeta=x/r_m$ and the term $V_b(\zeta)$ reads
$$
  V_b(\zeta)=A\exp(-\alpha\zeta
  +\beta\zeta^2 ) -\left [ \frac{C_6}{\zeta^6}
  + \frac{C_8}{\zeta^8} +
  \frac{C_{10}}{\zeta^{10} } \right ]F(\zeta)\,,
$$
$x$ is expressed in the same length units as $r_m$
({\AA} in the present case). The function $F(\zeta)$ is given by
$$
 F(\zeta)=\cases { \exp{ \left [ -\left( D/\zeta
 -1 \right) \right]^2},
    	& \mbox{if  $\zeta\leq D$}   \cr
1,  & \mbox{if  $\zeta >  D $}\,. }
$$
In addition to the term $V_b(\zeta)$ the LM2M2
potential~\cite{Aziz91} contains a term $V_a(\zeta)$,
\begin{equation}
\label{LM2M2}
  V(r)=\varepsilon \left\{V_b(\zeta)+V_a(\zeta)\right\},
\end{equation}
where
$$
V_a(\zeta)=\left\{\begin{array}{cl}
A_a\left\{\sin\left[\displaystyle
\frac{2\pi(\zeta-\zeta_1)}{\zeta_2-\zeta_1}-
\frac{\pi}{2}\right]+1\right\}, \quad & \zeta_1\leq\zeta\leq\zeta_2 \\
0, & \zeta\not\in[\zeta_1,\zeta_2]\,.
\end{array}\right.\,
$$
The parameters for the HFDHE2, HFD-B and LM2M2 potentials are
given in Table  \ref{tableAziz}.

The form of the theoretical He--He potential TTY is taken from
\cite{Tang95}. This potential reads
$$
 V(x)=A\,[V_{\rm ex}(x)+V_{\rm disp}(x)]\,
$$
where $x$ stands for the distance between the $^4$He atoms given in
atomic length units. (Following \cite{Tang95} in converting the
length units we used 1\,a.u.$=0.52917$\,{\AA}.) The
function $V_{\rm ex}$ has the form
$$
V_{\rm ex}(x)=D\,x^p\exp(-2\beta x)\,
$$
with $p=\displaystyle\frac{7}{2\beta}-1\,$, while the function
$V_{\rm disp}$ reads
$$
  V_{\rm disp}(x)=-\sum\limits_{n=3}^N C_{2n}\,f_{2n}(x)\,x^{-2n}\,.
$$
The coefficients $C_{2n}$ are calculated via the recurrency relation
$$
C_{2n}=\left(\displaystyle\frac{C_{2n-2}}{C_{2n-4}}\right)^3\,C_{2n-6}\,,
$$
and the functions $f_{2n}$ are given by
$$
   f_{2n}(x)=1-\exp(-bx)\,\sum\limits_{k=0}^{2n}
   \displaystyle\frac{(bx)^k}{k!}
$$
where
$$
   b(x)=2\beta-\left[\displaystyle\frac{7}{2\beta}-1\right]\,
   \displaystyle\frac{1}{x}\,.
$$
The parameters  of the TTY potential are given in Table \ref{tableTTY}.


\newpage



\begin{table}
\caption{Dimer energies $\epsilon_d$, inverse wave lengths
         $1/\varkappa^{(2)}$, and $^4$He$-$$^4$He scattering
         lengths $\ell_{\rm sc}^{(2)}$ for the potentials used.}
\label{tableDimerLen}
\begin{tabular}{|c|ccc||c|ccc|}
\hline
Potential & $E_d$ (mK) & $1/\varkappa^{(2)}$ (\AA)
& $\ell^{(2)}_{\rm sc}$ (\AA) &
Potential & $E_d$ (mK) & $1/\varkappa^{(2)}$ (\AA)
& $\ell^{(2)}_{\rm sc}$ (\AA) \\
\hline
 HFDHE2 & $-0.83012$ & 120.83 & $124.65$ &  LM2M2 & 1.30348 & 96.43 & 100.23 \\
 HFD-B  & $-1.68541$ &  84.80 & $ 88.50$ &  TTY   & 1.30962 & 96.20 & 100.01 \\
\hline
\end{tabular}
\end{table}


\begin{table}
\caption
{Ground state energy $E_t^{(0)}$ results for the helium trimer.
The (absolute) values of $E_t^{(0)}$ are given in K.
The grid parameters used were: $N_\theta=N_\rho=555$,
$\tau_0=0.2$, $\nu=3.6$,  and $\rho_{\rm max}=250$\,{\AA}.
}
\label{tableTrimerGS}
\begin{tabular}{|c|cccccc|ccccc|cc|}
\hline
 & \multicolumn{6}{c|}{Faddeev}
&\multicolumn{5}{c|}{Variational}&\multicolumn{2}{c|}{Adiabatic} \\
 & \multicolumn{6}{c|}{equations}
&\multicolumn{5}{c|}{methods}&\multicolumn{2}{c|}{approaches} \\
\cline{2-14}
Potential   & $l_{\rm max}$ &  This work & \cite{CGM} & \cite{Gloeckle}
& \cite{Nakai} & \cite{RoudnevYakovlev} &
\cite{Pandharip} & \cite{RickLynchDoll} &\cite{Barnett}& \cite{Lewerenz} &\cite{Guardiola}&
\phantom{}\cite{EsryLinGreene}&  \phantom{}\cite{Nielsen}\\
\hline \hline
 & 0 & $0.084^{a)}$ $0.0823$&  & $0.082$ & $0.092$ & & & & & & & $0.098$ & \\
\cline{2-6}
HFDHE2 & 2 &  $0.114^{a)}$ $0.1124$ & $0.107$ & $0.11$ & & & & & & & & &\\
\cline{2-4}
 & 4 & $0.1167$ & & & & 0.1171 &  $0.1173$  &  & & & & &\\
\hline \hline
 & 0 &$0.096^{a)}$ $0.0942$ & $0.096$ & & & & & & & & & &\\
\cline{2-4}
HFD-B& 2 & $0.131^{a)}$ $0.1277$ & $0.130$ & & & & & & & & & & \\
\cline{2-4}
 & 4  & $0.1325$ & & & & 0.1330 & & $0.1193$
                                      & $0.133$ &$0.131$& 0.129 & & \\
\hline
\hline
   & 0 &$0.0891$ & & & & & & & & & & $0.106$ &  \\
\cline{2-3}
LM2M2 & 2 & $0.1213$ & & & & & & & & & & & \\
\cline{2-3}
 & 4 & $0.1259$ &  & & & 0.1264 & & & & & & & $0.1252$\\
\hline \hline
    & 0 & $0.0890$  & & & & & & & & & & & \\
\cline{2-3}
TTY & 2 & $0.1212$  & & & & & & & & & & &\\
\cline{2-3}
    & 4 & $0.1258$  & & & & 0.1264 & & & & $0.126$ & & & \\
\hline
\end{tabular}
{\footnotesize\noindent$^{a)}$Results from~\cite{KMS-JPB} for a
grid with $N_\theta=N_\rho=275$ and $\rho_{\rm
max}=60$\,{\AA}.}\\
\end{table}


\begin{table}
\caption
{Excited state energy $E_t^{(1)}$ results for the helium trimer.
The (absolute) values of $E_t^{(1)}$ are given in mK.
The grid parameters used were: $N_\theta=N_\rho=805$,
$\tau_0=0.2$, $\nu_0=3.6$,  and $\rho_{\rm max}=300$\,{\AA}.
}
\label{tableTrimerExcS}
\begin{tabular}{|cccccccc|}
\hline
Potential   & $l_{\rm max}$ &  This work &  \cite{Gloeckle} & \cite{Nakai}
& \cite{RoudnevYakovlev} &\cite{EsryLinGreene} &\cite{Nielsen}\\
\hline \hline
            &  0  &  $1.5^{a)}$ $1.46$  & $1.46$  &  $1.04$ &  & $1.517$ & \\
\cline{2-5}
HFDHE2      &  2  &  $1.7^{a)}$ $1.65$  & $1.6$   &  & & & \\
\cline{2-4}
            &  4  &             $1.67$  &         &       & 1.665 & & \\
\hline \hline
            &  0  &  $2.5^{a)}$ $2.45$  &         &       &      &  &\\
\cline{2-3}
 HFD-B      &  2  &  $2.8^{a)}$ $2.71$  &         &       &      &  & \\
\cline{2-3}
            &  4  &             $2.74$  &         &       & 2.734 & & \\
\hline \hline
            &  0  &  $2.02$             &         &       &       & 2.118 & \\
\cline{2-3}
 LM2M2      &  2  &  $2.25$             &         &       &       &    & \\
\cline{2-3}
            &  4  &  $2.28$             &         &       & 2.271 &    & 2.269 \\
\hline \hline
            &  0  &  $2.02$    &         &  & & & \\
\cline{2-3}
 TTY        &  2  &  $2.25$    &         &  & & & \\
\cline{2-3}
            &  4  &  $2.28$    &         &  & 2.280 & & \\
\hline
\end{tabular}
{\footnotesize \noindent$^{a)}$Results from~\cite{KMS-JPB} for a
grid with $N_\theta=N_\rho=252$ and $\rho_{\rm
max}=250$\,{\AA}.}\\
\end{table}

\begin{table}
\caption
{Trimer excited-state energy $E_t^{(1)}$ (mK)
obtained with the TTY potential for various grids.
}
\label{tableExcGrids}
\begin{tabular}{|c|c|c|c|c|c|}
\hline
\phantom{AAAAAAA} &  $N_\theta=N_\rho=252$  &
                     $N_\theta=N_\rho=502$  &
                     $N_\theta=N_\rho=652$  &
                     $N_\theta=N_\rho=805$  &
                     $N_\theta=N_\rho=1005$  \\
$l_{\rm max}$  & $\rho_{\rm max}=250$\.\AA
               & $\rho_{\rm max}=300$\,\AA
               & $\rho_{\rm max}=300$\,\AA
               & $\rho_{\rm max}=300$\,\AA
               & $\rho_{\rm max}=300$\,\AA \\
\hline
 0     & $-2.108$ &  $-2.039$ & $-2.029$  & $-2.024$ & $-2.021$ \\
 2     & $-2.348$ &  $-2.273$ & $-2.258$  & $-2.253$ & $-2.248$ \\
\hline
\end{tabular}
\end{table}


\begin{table}
\caption{Estimations for $^4$He atom\,--\,$^4$He dimer
        scattering lengths $\ell_{\rm sc}$ and inverse wave numbers
        $\varkappa^{-1}$  corresponding to the
        excited-state energy $E_t^{(1)}$ for the HFD-B, LM2M2 and
        TTY potentials. The accuracy for the scattering lengths
        is within $\pm5$\,{\AA}. The grid parameters used for the
        calculation of $\ell_{\rm sc}$
        are:  $N_\theta= N_\rho=502$, $\tau_0=0.18$, $\nu=3.45$
        and $\rho_{\rm max}=460$\,{\AA}.
}
\label{tableTrimerLen}
\begin{tabular}{|c|ccc||c|ccc|}
\hline
Potential & $l_{\rm max}$ & $\ell_{\rm sc}$ (\AA) & $\varkappa^{-1}$ (\AA) &
Potential & $l_{\rm max}$ & $\ell_{\rm sc}$ (\AA) & $\varkappa^{-1}$ (\AA) \\
\hline
        & 0  & $170^{a)}$ $168$ & 109 &            & 0 & $168$ & 113 \\
 HFD-B  & 2  & $145^{a)}$ $138$ &  94 & LM2M2/TTY  & 2 & $134$ &  98 \\
        & 4  & $135$            &  93 &            & 4 & $131$ &  96 \\
\hline
\end{tabular}
{\footnotesize \noindent$^{a)}$Results from~\cite{KMS-JPB} for a
grid with $N_\theta=N_\rho=320$ and $\rho_{\rm
max}=400$\,{\AA}.}\\
\end{table}


\begin{table}
\caption
{Phase shift $\delta_0^{(l_{\rm max})}$  results (in degrees) for
the HFD-B potential for various c.m. energies $E$ (in mK).  The grid parameters
used are: $N_\theta=N_\rho=502$, $\tau_0=0.18$, $\nu=3.45$, and
$\rho_{\rm max}=$460\,{\AA}.
}
\label{tablePhaseHFDB}
\begin{tabular}{|c|c|c|c||c|c|c|c||c|c|c|c|}
\hline
$E$   & $\delta_0^{(0)}$ &  $\delta_0^{(2)}$  & $\delta_0^{(4)}$&
$E$   & $\delta_0^{(0)}$ &  $\delta_0^{(2)}$  & $\delta_0^{(4)}$ &
$E$   & $\delta_0^{(0)}$ &  $\delta_0^{(2)}$  & $\delta_0^{(4)}$ \\
\hline
$-1.68541 $ & $ 359.9 $  & $ 359.9$ & $ 359.9 $ &$-1.05    $ & $ 299.1 $  & $ 308.2$ & $ 309.2 $ &$0.95     $ & $ 262.4 $  & $ 272.1$ & $ 273.7 $ \\
$-1.68    $ & $ 352.6 $  & $ 353.9$ & $ 354.1 $ &$-0.8     $ & $ 290.8 $  & $ 300.4$ & $ 301.5 $ &$1.2      $ & $ 260.0 $  & $ 269.6$ & $ 270.7 $ \\
$-1.65    $ & $ 341.7 $  & $ 345.0$ & $ 345.4 $ &$-0.55    $ & $ 284.4 $  & $ 294.2$ & $ 295.4 $ &$1.45     $ & $ 257.8 $  & $ 267.3$ & $ 268.4 $ \\
$-1.60    $ & $ 330.8 $  & $ 337.7$ & $ 338.2 $ &$-0.3     $ & $ 279.3 $  & $ 289.3$ & $ 290.4 $ &$1.7      $ & $ 255.9 $  & $ 265.2$ & $ 266.3 $ \\
$-1.55    $ & $ 326.9 $  & $ 332.8$ & $ 333.5 $ &$-0.05    $ & $ 275.1 $  & $ 285.2$ & $ 286.3 $ &$1.95     $ & $ 254.1 $  & $ 263.4$ & $ 264.5 $ \\
$-1.50    $ & $ 322.4 $  & $ 329.0$ & $ 329.8 $ &$0.2      $ & $ 271.4 $  & $ 281.3$ & $ 282.5 $ &$2.2      $ & $ 252.5 $  & $ 261.7$ & $ 262.7 $ \\
$-1.40    $ & $ 315.4 $  & $ 323.0$ & $ 323.9 $ &$0.45     $ & $ 268.1 $  & $ 277.9$ & $ 279.0 $ &$2.45     $ & $ 251.0 $  & $ 260.1$ & $ 261.1 $ \\
$-1.30    $ & $ 309.9 $  & $ 318.1$ & $ 319.1 $ &$0.7      $ & $ 265.1 $  & $ 274.8$ & $ 276.0 $ &&&&\\
\hline
\end{tabular}
\end{table}


\begin{table}
\caption
{Phase shift $\delta_0^{(l_{\rm max})}$ results for the LM2M2 potential.  The
units and grid parameters used are the same as in Table
\ref{tablePhaseHFDB}.}
\label{tablePhaseLM2M2}
\begin{tabular}{|c|c|c||c|c|c||c|c|c|}
\hline
$E$   & $\delta_0^{(0)}$ &  $\delta_0^{(2)}$  &
$E$   & $\delta_0^{(0)}$ &  $\delta_0^{(2)}$  &
$E$   & $\delta_0^{(0)}$ &  $\delta_0^{(2)}$  \\
\hline
$-1.30348$ & $  359.8  $ & $ 359.9 $ &  $-0.8    $ & $  304.6  $ & $ 313.8 $ &  $ 0.95   $ & $  267.0  $ & $ 276.2 $  \\
$-1.3    $ & $  354.1  $ & $ 355.3 $ &  $-0.55   $ & $  295.2  $ & $ 304.8 $ &  $ 1.2    $ & $  264.1  $ & $ 273.2 $  \\
$-1.25   $ & $  337.9  $ & $ 342.3 $ &  $-0.3    $ & $  287.9  $ & $ 297.7 $ &  $ 1.45   $ & $  261.5  $ & $ 270.6 $  \\
$-1.20   $ & $  330.5  $ & $ 336.3 $ &  $-0.05   $ & $  282.3  $ & $ 292.2 $ &  $ 1.7    $ & $  259.2  $ & $ 268.1 $  \\
$-1.15   $ & $  325.2  $ & $ 332.0 $ &  $ 0.2    $ & $  277.7  $ & $ 287.4 $ &  $ 1.95   $ & $  257.1  $ & $ 266.0 $  \\
$-1.10   $ & $  321.1  $ & $ 328.5 $ &  $ 0.45   $ & $  273.7  $ & $ 283.2 $ &  $ 2.2    $ & $  255.3  $ & $ 264.0 $  \\
$-1.05   $ & $  317.6  $ & $ 325.5 $ &  $ 0.7    $ & $  270.1  $ & $ 279.5 $ &  $ 2.45   $ & $  253.6  $ & $ 262.3 $  \\
\hline
\end{tabular}
\end{table}


\begin{table}
\caption
{Phase shift $\delta_0^{(l_{\rm max})}$ results  for the TTY
potential.  The units and grid parameters used are the same as in Table
\ref{tablePhaseHFDB}.}
\label{tablePhaseTTY}
\begin{tabular}{|c|c|c|c||c|c|c|c||c|c|c|c|}
\hline
$E$   & $\delta_0^{(0)}$ &  $\delta_0^{(2)}$  & $\delta_0^{(4)}$&
$E$   & $\delta_0^{(0)}$ &  $\delta_0^{(2)}$  & $\delta_0^{(4)}$ &
$E$   & $\delta_0^{(0)}$ &  $\delta_0^{(2)}$  & $\delta_0^{(4)}$ \\
\hline
$-1.30961 $ & $ 359.7 $  & $ 359.8$ & $359.8$ & $-0.8     $ & $ 304.3 $  & $ 313.5$ & $314.6 $ &  $ 0.95    $ & $ 266.8 $  & $ 276.1$ & $277.2 $ \\
$-1.308   $ & $ 355.9 $  & $ 356.8$ & $356.9$ & $-0.55    $ & $ 295.0 $  & $ 304.6$ & $305.7 $ &  $ 1.2     $ & $ 264.0 $  & $ 273.1$ & $274.2 $ \\
$-1.3     $ & $ 350.2 $  & $ 352.1$ & $352.4$ & $-0.3     $ & $ 287.7 $  & $ 297.5$ & $298.7 $ &  $ 1.45    $ & $ 261.4 $  & $ 270.5$ & $271.5 $ \\
$-1.25    $ & $ 336.8 $  & $ 341.4$ & $341.9$ & $-0.05    $ & $ 282.0 $  & $ 292.0$ & $293.2 $ &  $ 1.7     $ & $ 259.1 $  & $ 268.1$ & $269.1 $ \\
$-1.2     $ & $ 329.7 $  & $ 335.7$ & $336.4$ & $ 0.2     $ & $ 277.5 $  & $ 287.3$ & $288.4 $ &  $ 1.95    $ & $ 257.0 $  & $ 265.9$ & $266.9 $ \\
$-1.10    $ & $ 320.5 $  & $ 328.1$ & $329.0$ & $ 0.45    $ & $ 273.5 $  & $ 283.1$ & $284.2 $ &  $ 2.2     $ & $ 255.0 $  & $ 263.9$ & $265.0 $ \\
$-1.05    $ & $ 317.1 $  & $ 325.1$ & $326.1$ & $ 0.7     $ & $ 270.0 $  & $ 279.4$ & $280.5 $ &  $ 2.45    $ & $ 253.5 $  & $ 262.2$ & $263.2 $ \\
\hline
\end{tabular}
\end{table}


\begin{table}
\caption{The parameters for the  $^4$He$-$$^4$He Aziz and co-workers
          potentials used.}
\label{tableAziz}
\begin{tabular}{llll}
\hline
Parameter & HFDHE2 \cite{Aziz79} & HFD-B \cite{Aziz87} &
LM2M2 \cite{Aziz91} \\\hline
   $\varepsilon$ (K)      &    10.8       &  10.948    &   10.97 \\
   $ r_m $ (\AA)          &   2.9673      &  2.963     &   2.9695  \\
   $A$                    &   544850.4    &  184431.01 &  189635.353 \\
   $\alpha$               & 13.353384     & 10.43329537 & 10.70203539\\
   $\beta$                &     0         & $-2.27965105$& -1.90740649 \\
   $C_6$                  & 1.3732412     & 1.36745214 & 1.34687065  \\
   $C_8$                  & 0.4253785     & 0.42123807 & 0.41308398 \\
   $C_{10}$               & 0.178100      & 0.17473318 & 0.17060159 \\
   $D$                    & 1.241314      & 1.4826     & 1.4088 \\
   $A_a$                  & $-$           & $-$        & 0.0026  \\
   $\zeta_1$              & $-$           & $-$        & 1.003535949\\
   $\zeta_2$              & $-$           & $-$        & 1.454790369\\
\hline
\end{tabular}
\end{table}


\begin{table}
\caption{The parameters for the  $^4$He$-$$^4$He TTY
          potential used.}
\label{tableTTY}
\begin{tabular}{|ll||rl|}
\hline
   $A$ (K) &  $315766.2067^{a)}$                       &  $C_6$ & $1.461$  \\
$\beta$ $\bigl($(a.u.)$^{-1}$$\bigr)$ & $1.3443$  &$C_8$&$14.11$\\
   $D$        &   $7.449$                              &  $C_{10}$   &   $183.5$  \\
   $N$        &   $12$      &             &            \\
\hline
\end{tabular}
{\footnotesize
\noindent $^{a)}$The value of $A$ was obtained from the data
presented in \cite{Tang95} using, for converting the energy units,
the factor 1\,K$=3.1669\times 10^{-6}$\,a.u.}
\end{table}



\end{document}